# On the Magnetism of the Normal State in MgB$_2$

**S. Reich**[*] and **G. Leitus**

Department of Materials & Interfaces
The Weizmann Institute of Science, Rehovot 76100, Israel

**I. Felner**

Racah Institute of Physics, The Hebrew University
Jerusalem 91904, Israel

**ABSTRACT**

The experimentally observed ferromagnetism in MgB$_2$ in the normal state is attributed to micro phase separated inclusions of iron. This magnetic character is also observed when the iron content of the samples is reduced below 20 µg/g, however in these samples the diamagnetism of MgB$_2$ is apparent and is measured. It is found experimentally that the diamagnetic susceptibility at room temperature of B element, MgB$_2$ and MgB$_4$ is close to the ratio 1:2:4, suggesting that the diamagnetism in these borides is confined to the boron atoms. This observation supports a picture in which the two electrons of Mg are donated to B in MgB$_2$

[*] cpreich@wicc.weizmann.ac.il

It was observed by researchers (1) that commercial as well as in house prepared $MgB_2$ samples exhibit ferromagnetic character in the normal state. This behaviour was attributed to magnetic impurities, mainly of iron and cobalt in the concentration range of few hundred to few thousands µg/g sample.

We prepared samples of $MgB_2$ and $MgB_4$ by standard procedures (2) using 99.99% purity magnesium and boron. To our surprise the $MgB_4$ samples were diamagnetic, see Fig. 1, while the $MgB_2$ samples were indeed ferromagnetic in the normal state. Our first idea was that in the preparation of $MgB_2$ the formation of FeB, a ferromagnetic compound (3), is favored. We prepared FeB and measured its critical temperature, $T_c$=625 K, and its magnetic saturation at 300K, $\sigma_s$= 61 emu/g, see Fig. 2. By measuring the saturation magnetization of $MgB_2$ vs. temperature above 300K we found however that at 625 K, the critical temperature of FeB, it decreased only by 26% of its room temperature value. This suggests that the critical temperature for the magnetic impurity in the $MgB_2$ sample is substantially higher ~ 1000 K. This observation suggests iron metal, $T_c$=1044 K, as the main impurity as Co and Ni were essentially eliminated completely in very pure samples described below. The question then, is why we do not observe similar magnetic signal in $MgB_4$ samples, as the two compounds were prepared from the same starting materials.

In view of the above we prepared $MgB_2$ from very pure elements; B (99,999%) and clean Mg with an iron content below 20 µg/g, see table I. In this table we also present the values for magnetic contaminants in a typical commercial $MgB_2$ material. The magnetization loops for the commercial and our "pure" samples at RT are presented in Fig. 3. We observe a drastic reduction in the magnetization, a reduction in coercive force, from $H_c \sim 450$ Oe to $\sim 30$ Oe, and a diamagnetic trend for the magnetization response for $|H| > 2500$ Oe. The reduction in coercivity is probably due to essentially total elimination of Co and Ni, the reduction in the amplitude of magnetization is due to a very small iron content <20 µg/g vs. 570 µg/g in the commercial sample. The diamagnetic response at high fields is due to the presence of unreacted boron, $MgB_4$, and the dominant $MgB_2$ phase.

By X-ray analysis on the "pure" material we obtained the following volume composition: B - 8%, $MgB_4$ - 18 %, and $MgB_2$ - 74 %. Knowing these values and measuring the diamagnetic response for pure boron, 99.999% purity, and $MgB_4$, see Fig. 1 we were able to correct for the diamagnetic contributions of these phases. The as measured and the corrected magnetic loops for the "pure" samples at 45K and 300K are shown in Fig. 4. The corrected curves yield the diamagnetic susceptibility within an accuracy of 10 % for pristine $MgB_2$ at 45 and 300K, see table II. Note that the susceptibilities at R.T. for B, $MgB_2$ and $MgB_4$ are close to the ratio 1:2:4 showing that all the electronic diamagnetic action in those materials is confined to the boron atoms. This observation supports a picture in which the two electrons of Mg are donated to B in $MgB_2$ compound (4-5).

The fact that the "pure" $MgB_2$ still shows a ferromagnetic contribution to the signal strongly suggests that iron is micro phase separated in this compound.

On the other hand at small concentration, iron dissolves well in MgB$_4$ and does not mask the diamagnetic response of this material.

**Acknowledgements**:

The authors are grateful to M. Kazir from the Dead Sea Magnesium Ltd. Beer-Sheva for supplying the "clean" Mg material.

This research was supported by the BSF (1998), by the Klachky Foundation and by the Israel Science Foundation (2000).

**Table I**

**Concentration in μg/g as obtained by ICP**

|  | Fe | Co | Ni |
|---|---|---|---|
| Mg | <20 | <2 | <5 |
| B | 0.16 | - | - |
| $MgB_2$ from "Alfa" | 570 | 65 | 800 |

**Table II**

**Magnetic susceptibility in emu/g mole Oe x $10^6$**

|  | B | $MgB_2$ | $MgB_4$ |
|---|---|---|---|
| T=45 K | -9.01 | -13.4 | -26.1 |
| T=300 K | -8.27 | -19.5 | -40.9 |

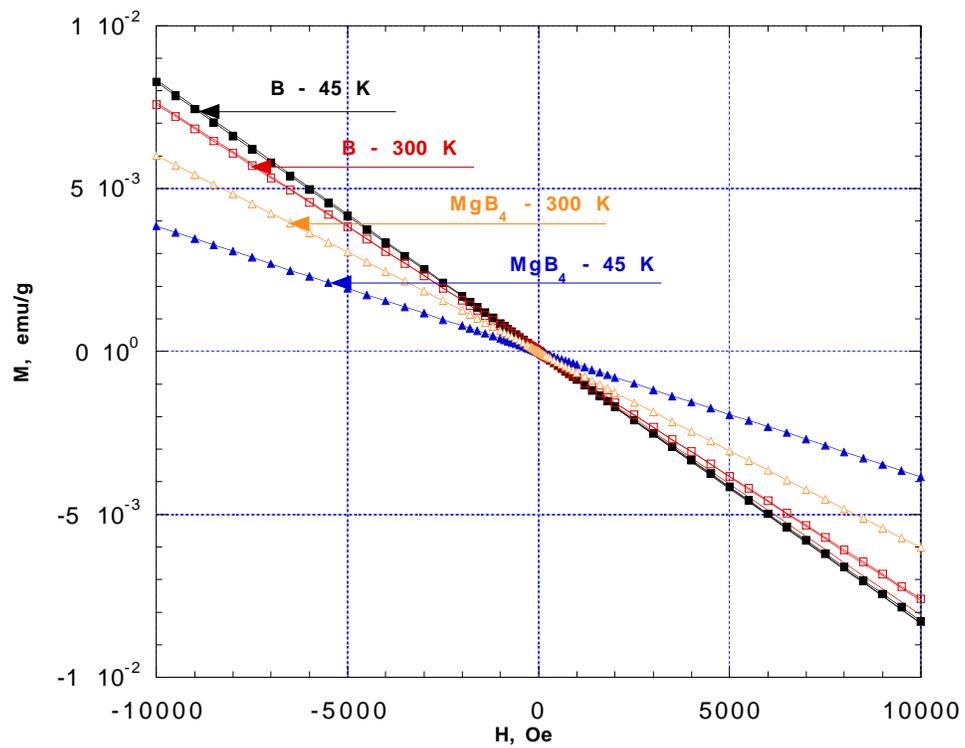

**Fig. 1**: Magnetization vs. field for B, 99.999 % purity, and for $MgB_4$ at 45 K and 300 K.

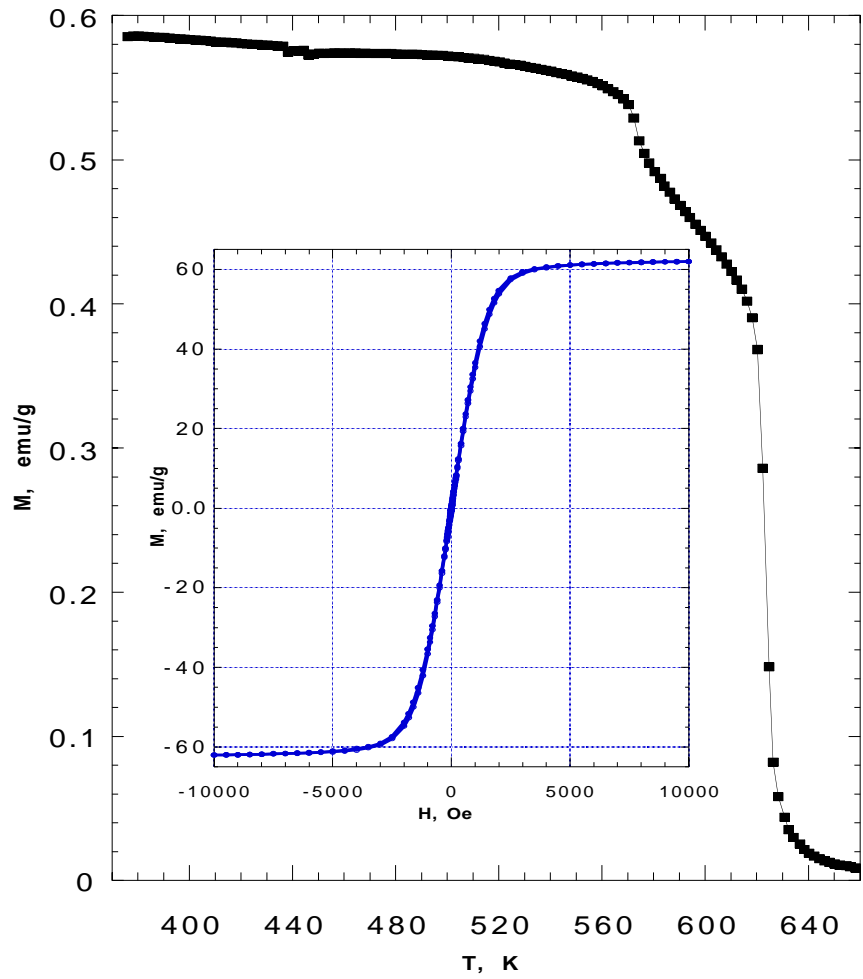

**Fig. 2**: Magnetization vs. temperature for FeB measured at 9 Oe. Inlay; magnetization loop at 300 K.

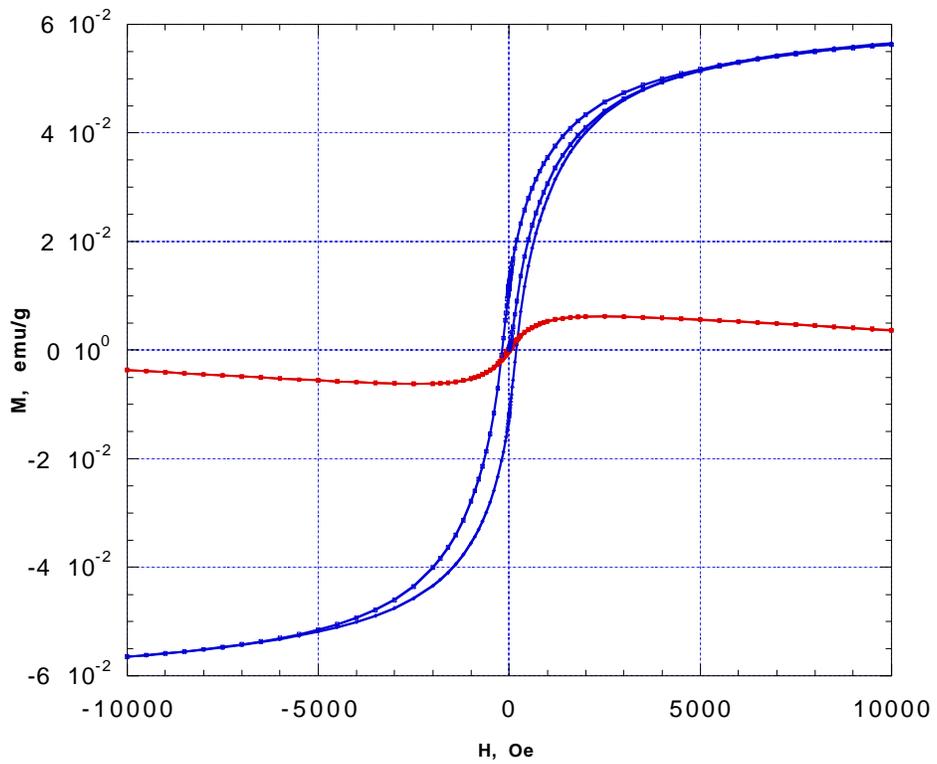

**Fig**. 3: Magnetization loops at 300 K for commercial $MgB_2$ from "Alfa" (large loop) and for a $MgB_2$ sample with iron content below 20 µg/g.

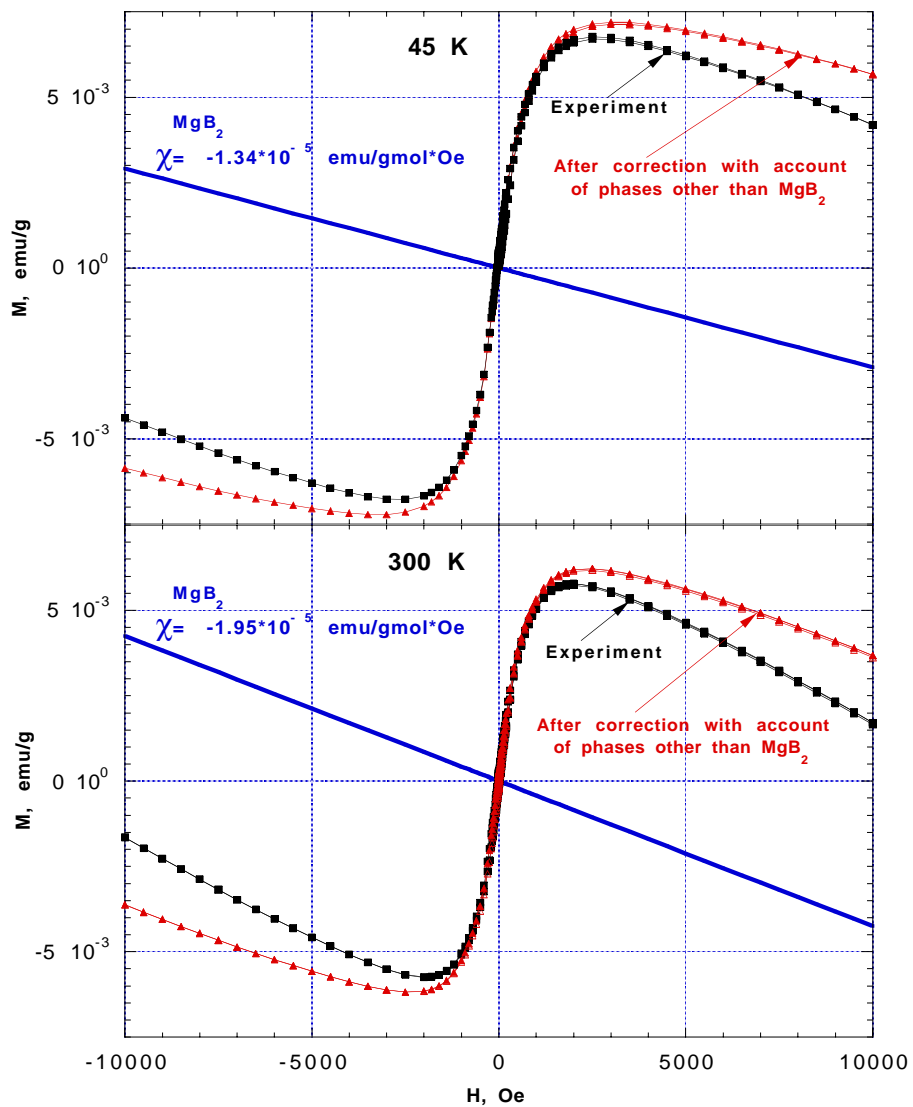

**Fig. 4**: Magnetization loops measured at 45 K and 300 K for $MgB_2$ sample with an iron content below 20 µg/g are presented. The experimental loops as well as the corrected loops for B and $MgB_4$ phases are shown. The diamagnetic response calculated for pristine $MgB_2$ is depicted by the straight line.